\documentclass[aps,prl,reprint,nofootinbib,twocolumn,superscriptaddress,showpacs,showkeys,longbibliography]{revtex4-1}
\usepackage{eurosym}
\usepackage{amsmath,amssymb,amstext}
\usepackage[usenames,dvipsnames]{color}
\usepackage{graphicx}
\usepackage{braket}
\usepackage{natbib}
\usepackage{comment}
\usepackage{dcolumn}
\usepackage[english]{babel}
\usepackage{wasysym}
\usepackage[colorlinks,bookmarks=false,citecolor=blue,linkcolor=red,urlcolor=blue]{hyperref}
\begin{document}
\title{  Interaction induced bi-skin effect in an exciton-polariton system}
\author{Xingran Xu}
\email{thoexxr@hotmail.com}
\affiliation{Division of Physics and Applied Physics, School of Physical and Mathematical Sciences, Nanyang Technological University, Singapore 637371, Singapore}
\author{Huawen Xu}
\affiliation{Division of Physics and Applied Physics, School of Physical and Mathematical Sciences, Nanyang Technological University, Singapore 637371, Singapore}
\author{S. Mandal}
\affiliation{Division of Physics and Applied Physics, School of Physical and Mathematical Sciences, Nanyang Technological University, Singapore 637371, Singapore}
\author{R. Banerjee}
\affiliation{Division of Physics and Applied Physics, School of Physical and Mathematical Sciences, Nanyang Technological University, Singapore 637371, Singapore}
\author{Sanjib Ghosh}
\affiliation{Division of Physics and Applied Physics, School of Physical and Mathematical Sciences, Nanyang Technological University, Singapore 637371, Singapore}
\author{T.C.H. Liew}
\email{TimothyLiew@ntu.edu.sg}
\affiliation{Division of Physics and Applied Physics, School of Physical and Mathematical Sciences, Nanyang Technological University, Singapore 637371, Singapore}
\affiliation{MajuLab, International Joint Research Unit UMI 3654, CNRS, Universit\'e C\^ote d'Azur, Sorbonne Universit\'e, National University of Singapore, Nanyang Technological University, Singapore}

\date{\today}

\begin{abstract}
The non-Hermitian skin effect can be realized through asymmetric hopping between forward and backward directions, where all the modes of the system are localized at one edge of a finite 1D lattice. However, achieving such an asymmetric hopping in optical systems is far from trivial. Here we show theoretically that  in a finite chain of 1D exciton-polariton micropillars with symmetric hopping, the inherent non-linearity of the system can exhibit a bi-skin effect, where the modes of the system are localized at the two edges of the system. To show the topological origin of such modes, we calculate the winding number. 
\end{abstract}
\maketitle

\emph{Introduction.---}Exciton-polaritons are bosonic quasiparticles arising from the strong coupling between  excitons and photons in microcavities~\cite{Carusotto2013,Rev1,Rev2,Rev3}. The combination of photons and excitons allows polaritons to possess the characteristics of both its root parts, for example, low effective mass from the photonic part and strong nonlinearity from the excitonic part. Moreover, due to the photons escaping from the microcavity, exciton-polaritons are noticeably an open quantum (non-Hermitian) system, which requires additional pumping to maintain a steady state in the system~\cite{Savvidis2001}. This character makes the exciton-polariton system an ideal platform to study non-Hermitian transitions with gain and loss~\cite{Nalitov2019,xu2020}. Although various research has been done in the Hermitian regime to realize polariton topological phases with the interplay between Zeeman shift resulting from the application of magnetic field and the transverse electric-transverse magnetic (TE-TM) splitting of the photonic modes~\cite{Bardyn2015,Nalitov2015,Li2018,Gulevich2016,Klembt2018}, accounting for nonlinearity~\cite{Bardyn2016, Sigurdsson2017}, and using the polarization splitting of elliptical micropillars~\cite{Banerjee2018}, several works have been reported in recent years to explore the non-Hermitian physics in the exciton-polariton system. Exceptional points (EPs) in the exciton-polariton system have been realized in  Refs.~\cite{Gao_2015, Gao2018a,Gao_2018}. Recently, the measurement of non-Hermitian topological invariants~\cite{Su2020}, topological end mode lasing~\cite{Comaron2020}, and non-reciprocal polaritons~\cite{Mandal2020} have been reported.

Non-Hermitian physics has attracted a tremendous attention in recent years due to the discovery of the non-Hermitian skin effect, where {\it all} the modes are localized to one end of a lattice~ \cite{Bandres_2018,Shen2018,Kawabata_2019,Longhi2019,Ghatak_2019,hou2020twodimensional,Weidemann311,Lee_2019,Hofmann2020,Zhou2019,Liu2020,Luo2019,Liutao2020,Topoorg,BBC1,skinmode1,Li_2020,Xiao_2020,song2019,song2019R,Lu2020,Lee2019,fu2020nonhermitian,Okugawa2020}. This is different to the case of Hermitian systems, where modes within a topological bandgap are localized at the edges of a finite sample due to bulk-boundary correspondence. Due to non-Hermiticity, the Bloch theory does not even hold approximately for finite sized non-Hermitian systems and one must turn to the generalized Brillouin zone (GBZ) theory based on a complex momentum to explain the non-Hermitian skin effect~\cite{Topoorg,Kunst2018,Kawabata2019,Masahito2018,yao2018,Lieu2018}. To realize the skin effect, the simplest model is the so-called Hatano-Nelson(HN) model \cite{HNmodel1,HNmodel2} without disorder where the hopping in a lattice is different in different directions.

Along with the same idea of the HN model, the Su-Schrieffer-Heeger(SSH) model with asymmetric hopping has been proposed or realized in many systems like waveguides~\cite{Zeuner2015,Yokomizo_2020}, photonic lattices~\cite{El_Ganainy_2019,Xiao_2020}, circuits~\cite{Helbig_2020} etc. The non-Hermitian topological bulk-boundary correspondence can be obtained by using the GBZ theory and by solving the boundary equations with complex momentum~\cite{Liusecond,yao2018,Yao_2018,Topoorg,BBC1,Jiangping2020,Imura2019,skinmode1}. Due to the chiral symmetry, the system will have an energy pair of ($E$,$-E$) and they will collapse at zero energy modes. The winding number can be well defined by the calculation with the periodic boundary energies with the GBZ.  Meanwhile, more and more work is going to explain the topological origin of the skin modes and the GBZ is not necessary for line gap non-Hermitian Hamiltonians where the basis can be enlarged to construct a Hermitian Hamiltonian~\cite{skinmode1,Liu_2020,Topoorg,Masahito2018,Lee_2019,Lee_20192}. Although the winding in the whole Bloch region is still zero for zero energy pairs, the system can still have a topological non-trivial phase if energy pairs can be made to collapse at non-zero (complex) energy.

\begin{figure}[h!]
\centering
\includegraphics[width=0.4\textwidth]{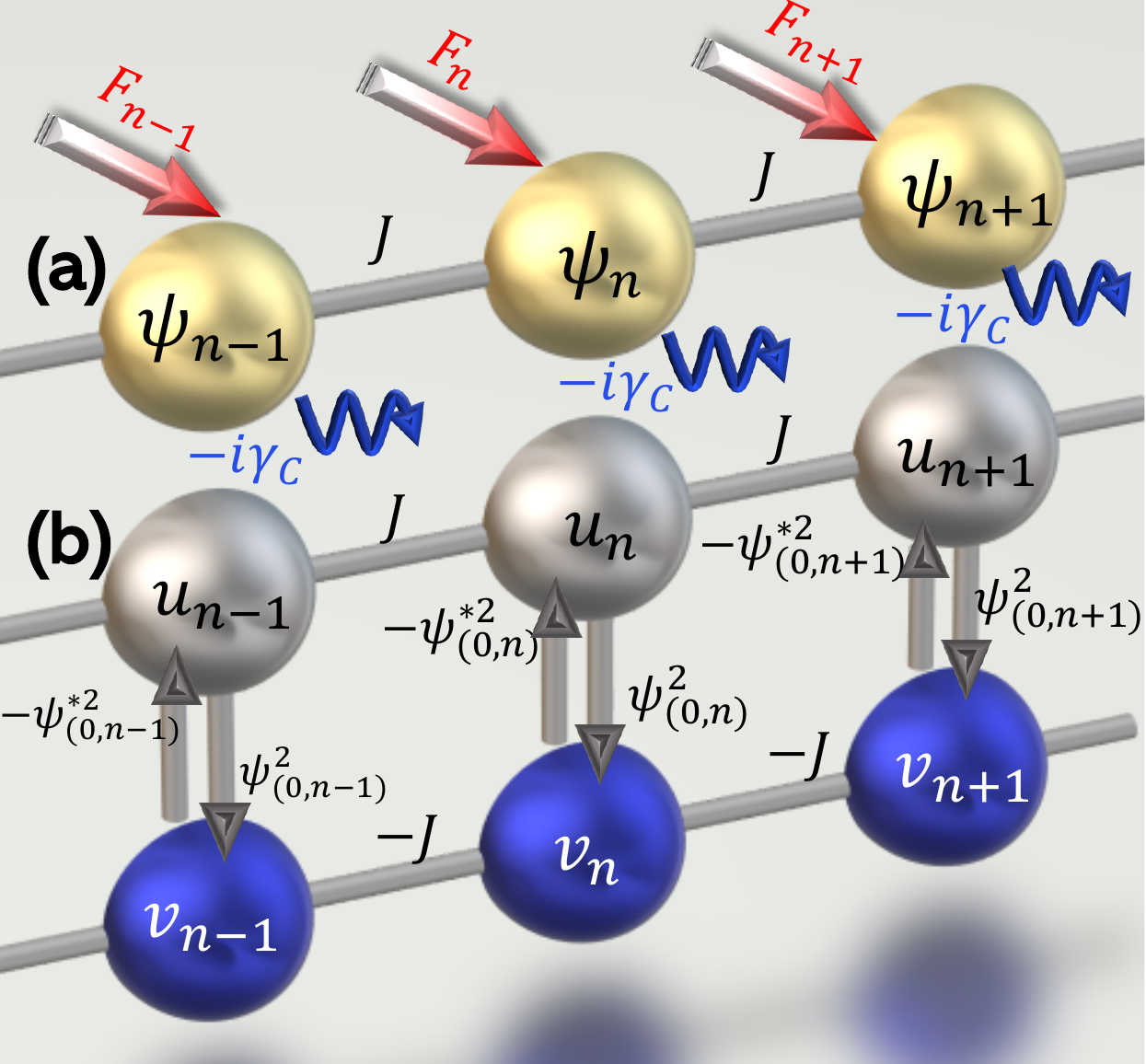}\\
\caption{Scheme of polariton lattice (a) and the related Bougoliubov lattice (b)  with $u_n$ and $v_n$ representing the different Bogoliubov fluctuation amplitudes on each site.}\label{scheme}   
\end{figure}

In this letter, we use the particle-hole symmetry of the fluctuation theory of exciton-polariton systems to realize ``bi-skin" modes where all the wavefunctions are localized at both sides of a one-dimensional lattice. Being compatible with recent experiments, we consider a non-equilibrium polariton condensate corresponding to a plane-wave with non-zero momentum. By the Bogoliubov-de-Gennes(BdG) transformation \cite{xu2017,Rimi2020}, the fluctuations of the condensate will have two modes with particle-hole symmetry. By considering a suitable BdG lattice by tight-binding theory, the bi-skin effect will appear with symmetric hopping  between sites. To classify the symmetry and the topologies of the system, we calculate the winding number of the system without assuming the GBZ theory. Because of the particle-hole symmetry~\cite{Okuma2019}, the energy spectrum will not collapse at zero modes but some purely imaginary energies.  The different topological phases can be well defined by the winding of the EPs and the localization of the wavefunction. Finally, we analyze the stability of the system and calculate the energy dispersion, which can be observed via photoluminescence~\cite{Utsunomiya_2008,Pieczarka2015} or four-wave mixing spectroscopy~\cite{Kohnle2011}.

\emph{Model.---}We consider a simple chain with symmetric hopping $J$  between nearest neighbours, formed by exciton-polariton micropillars, as shown in Fig.~\ref{scheme}(a). Such chains have been realized experimentally~\cite{Winkler2016}. Typically the behavior of a polariton condensate is studied, which itself has no topological features in such a system. The system can be described
by the driven-dissipative nonlinear Gross-Pitaevskii (GP) equation
\begin{eqnarray}
i\hbar \frac{\partial \psi_n}{\partial t}&=&\left(\Delta-i\gamma_C\right)\psi_n+ J(\psi_{n-1} + \psi_{n+1})\nonumber \\
&+&g\left|\psi_n\right|^2\psi_n+F_n,\label{HM0}
\end{eqnarray}
Here, $\Delta=E_0-\hbar\omega_p$ is the onsite detuning  between the bare
polariton mode having energy $E_0$ and a coherent drive $F_n$ with frequency $\omega_p$. $g$ represents the nonlinear interaction strength and $\gamma_C$ is the effective dissipation. In order to find the optimum parameter range we move to the dimensionless units by making the following substitutions:  $t\rightarrow t\hbar/\gamma_C$, $J\rightarrow J/\gamma_C$, $\Delta\rightarrow\Delta/\gamma_C$, $\psi_n\rightarrow\psi_n\sqrt{\gamma_C/g}$, and $F_n\rightarrow F_n\sqrt{g/\gamma_C^3}$. One of the main ingredients of our scheme is a plane-wave like stationary state $\psi_{0,n}=Ae^{ik_p n}$  (as shown in Fig.~\ref{wf}(a) ) realized by spatially modulating a coherent field~\cite{Ohadi2018,Alyatkin2020,Pickup_2020,Carusotto2004} and the appropriate choice of $F_n$ is obtained by setting the time derivative in Eq. (\ref{HM0}) to zero:
\begin{equation}
F_n=\left[-\Delta+i -2J\cos(k_p)-\left| A\right|^2 \right]Ae^{ik_pn},\label{Fn}
\end{equation} 
where $A$ is the square root of the density of the condensate and $k_p$ is the pumping momentum and $n\neq1$ or  $N$.  Because of the boundary, we should set the  first and the last site pumping by replacing $2J\cos(k_p)$ in Eq. (\ref{Fn}) with $Je^{ik_p}$ and $Je^{-ik_p}$, respectively. 

\begin{figure}[h!]
\centering
\includegraphics[width=0.48\textwidth]{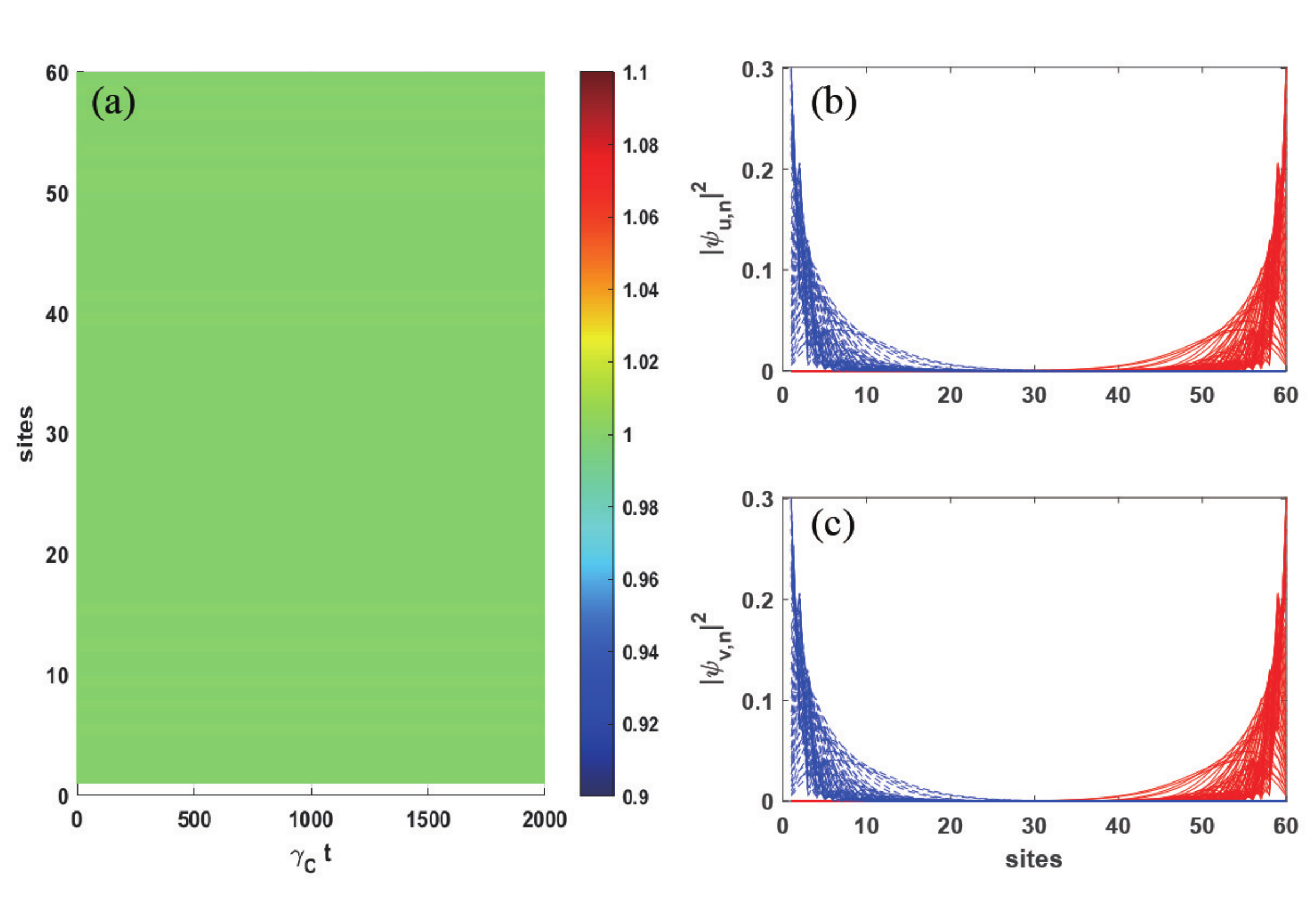}\\
\caption{The time evolution of the absolute square of the wavefunction $|\psi_n|^2$ (a)  described by Eq. (\ref{HM0}) and the spatial profiles of Bogoliubov modes $u_n$ and $v_n$ are plotted in (b) and (c), which are  obtained by Eqs.(\ref{Enu})- (\ref{Env}). The parameters are used $\Delta'/\gamma_C$=-0.2, $J$=0.4, $A$=1, and $k_p$=$\pi/6$.}\label{wf}
\end{figure}

To show the bi-skin effect we consider the fluctuations, by introducing the fluctuation amplitudes $u_n$ and $v_n$, and writing $\psi_n=\psi_{(0,n)}+u_ne^{iE t/\hbar}+v_n^*e^{-iE^*t/\hbar}$. The BdG modes as shown in Fig. \ref{scheme}(b) are derived by substituting $\psi_n$ into the mean-field GP equation, which yields the time-independent equations:
\begin{eqnarray}
E u_n&=&\left( \Delta' -i\right) u_n+J\left(u_{n+1}+u_{n-1}\right)+\psi^2_{(0,n)}v_n,\label{Enu}\\
E v_n&=&-\left(i +\Delta' \right)v_n-J\left(v_{n+1}+v_{n-1}\right)-\psi^{*2}_{(0,n)}u_n.\label{Env}
\end{eqnarray}
Here, $\Delta'=\Delta+2\left|\psi_{(0,n)}\right|^2$ is an effective detuning. The BdG transformation ensures that $u_n$ and $v_n$ modes have the particle-hole symmetry, so the hopping energies for $u_n$ and $v_n$ are $\pm J$ and the inter hopping energies in one unit cell are $\pm \psi_{(0,n)}^{(*)2}$. We need to mention that the hopping strength between different pairs of neighboring sites have the same magnitude but differ in their phases because of the symmetry. The Bogoliubov matrix $H_n$ with $N$ sites obtained by the Eqs. (\ref{Enu})-(\ref{Env})  can be diagonalized by $V^{-1}\hat{H_n}V=\text{diag}\left(
E_{N\times N}, -E_{N\times N}^{*}\right)$ with $V$ the matrix formed by eigenvectors, meanwhile, $S_x V^{-1}\hat{H_n}V S_x=\text{diag}\left(
E_{N\times N}, -E_{N\times N}^{*}\right)$ where $S_x V$ can also be the eigenvectors of the system~\cite{xu2017,Rimi2020}, here, $S_{x,y,z}=\hat{I}_{N,N}\otimes\sigma_{x,y,z}$ and $\sigma_{x,y,z}$ are the Pauli matrices. On the other hand,  the system has pseudo-Hermiticity \cite{yokomizo2020nonbloch,Kawabata2019} defined as $S_z \hat{H_n} S_z=\hat{H_n}^\dagger$. The particle-hole symmetry and  pseudo-Hermiticity guarantee that the system can have a 1D topological transition~\cite{Topoorg,yokomizo2020nonbloch} and the energy spectrum in real space is highly symmetrical.

\emph{The bi-skin modes.---}The stationary states of the bulk polaritons can have non-zero momentum induced by the specific $F_n$ pumping laser  and the nonlinearity makes the fluctuations modes $u_n$ and $v_n$ have a momentum shift. The momentum shift will just influence the phases of the energy spectrum under the periodic boundary condition (PBC), however, the energy spectrum under the open boundary condition (OBC) will completely change for the fluctuations at the edge can not go further and then become localized.

The model is equivalent to say that the hopping energies between the $ u_n$ and $v_n$ sites are different, but we stress that this is an effect induced by nonlinear interactions, where the fluctuations are affected by the non-zero momentum of the considered polariton condensate (stationary state). The underlying physical system is still none other than a regular polariton lattice with symmetric Hermitian hopping between lattice sites. As is shown in Figs. \ref{wf}(b) and (c), all eigenstates for $u_n$ and $v_n$ sites are localized at both sides. The energy spectrum of $u_n$ sites are $(E,E^*)-i$ pairs, while the energy spectrum of $v_n$ sites are $(-E^*,-E)-i$ pairs.   All eigenvectors of $E-i$ and $-E^*-i$ will be localized at one side and wavefunctions of the $E^*-i$ and $-E-i$ will be localized at the opposite side. Remarkably, the localization here does not mean the polariton's density localization but the fluctuations on the top of the polaritons.

\begin{figure}[h!]
\centering
\includegraphics[width=0.5\textwidth]{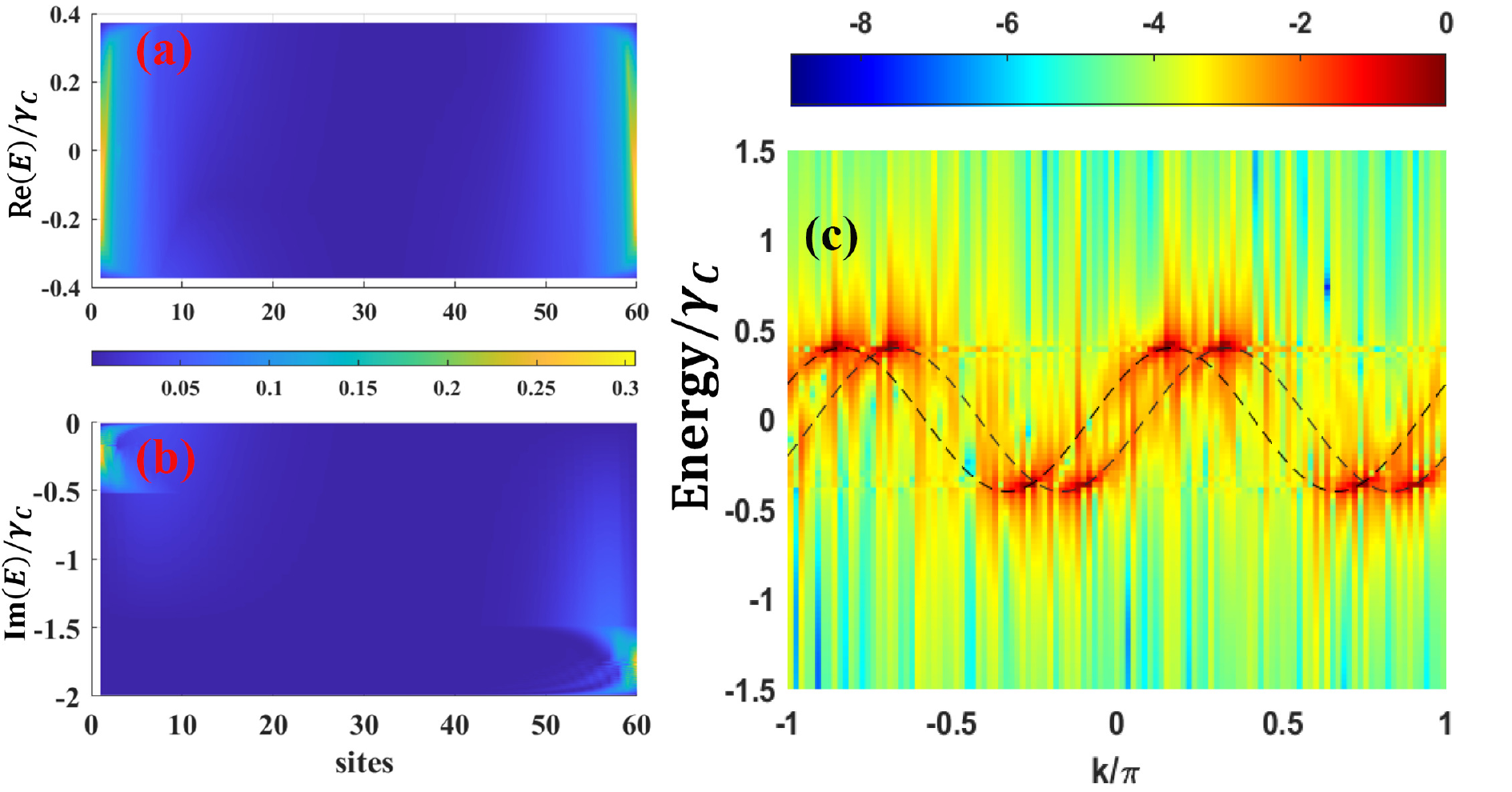}\\
\caption{Profiles of the real (a) and imaginary (b) eigenstates as a function of $u_n$ sites and eigenenergies. (c) Numerical results of the energy spectrums solved by Eqs. (\ref{Enu})-(\ref{Env}) with continuous pumping and the dashed lines are the fitting curves of the energy dispersion. Parameters are the same as in Fig. \ref{wf}. }\label{local}
\end{figure}

\emph{Stability.--- }Our model is based on the fluctuations of exciton-polaritons with a non-zero momentum stationary state. Because of the symmetry of the system, the imaginary parts of the eigenenergies can be beyond zero. However, with sufficient dissipation, there will be a total energy shift $i\gamma_C$ that keeps all imaginary energy components below zero.

\begin{figure}[h!]
\centering
\includegraphics[width=0.5\textwidth]{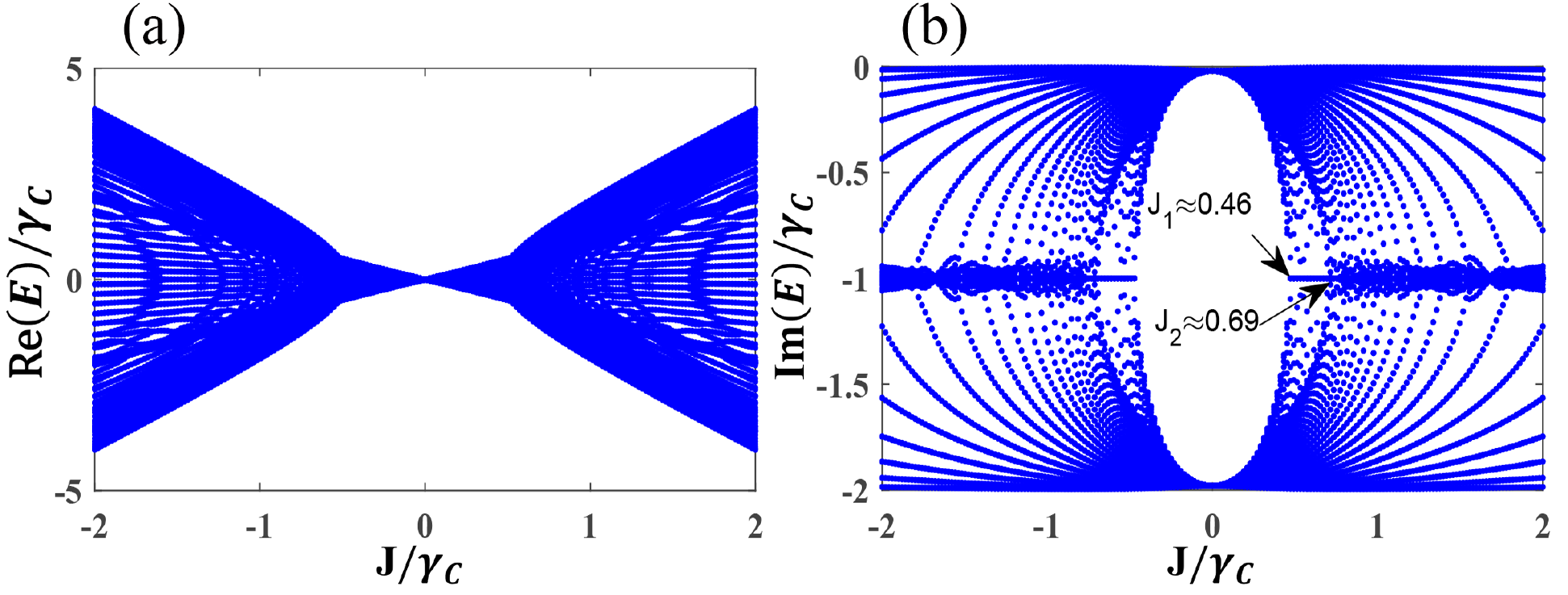}\\
\caption{ The real (a) and the imaginary (b) energy spectrum  obtained by Eqs.(\ref{Enu})-(\ref{Env}). Other parameters are the same as in Fig. \ref{wf}. }\label{real_BZ}
\end{figure}

Figs \ref{local}(a) and (b) show the distribution of intensity of the real and imaginary parts of the fluctuation eigenstates in real space. All fluctuation eigenstates are localized at the edges of the system, while we find that those with least negative imaginary part are localized on the left side.

Because the imaginary parts of the energy spectrum are larger than the real parts, a fundamental problem is whether the spectrum can be observed experimentally. To prove this, we numerically calculate the BdG lattice's energy dispersion in Fig. \ref{local}(c). We take $u_n$ and $v_n$ as perturbations with small random number compared with $\psi_{(0,n)}$ and solve the BdG equations. The dashed fitting curves have the same energy dispersion with diagonalization results but doubly shrinks the period for two inner sites here. The imaginary parts of the energy dispersion are different  for the  linewidth of the energy dispersion are different as a function of momentum. Remarkably, we solve the Eqs. (\ref{Enu})-(\ref{Env}) without coherent pumping here, which reveals our model is universal and can be extended to other systems.

\emph{Non-Hermitian topological invariants--}   The effective Hamiltonian can be written into the momentum space  with use of the Fourier transformation $u_n \rightarrow \frac{1}{\sqrt{N}}\sum_k \hat{a}_k e^{ikn}$ and $v_n\rightarrow \frac{1}{\sqrt{N}}\sum_k \hat{b}_k e^{ikn}$ with $\hat{a}_k$ and $\hat{b}_k$ the annihilation operators for the momentum $k$, giving
\begin{equation}
H_{k}=\left(\begin{array}{cc}
\Delta'+2J\cos\left(k\right) & A^{2}\\
-A^{2} & -\Delta'-2J\cos\left(k-2k_{p}\right)
\end{array}\right)-i\hat{I},\label{Hk2}
\end{equation}
where $\hat{I}$ is the $2\times2$ identity matrix and the periodic bounary condition is considered and the basis vector is $(\hat{a}_k,\hat{b}_{k-2k_p})$. The imaginary energy shift $i$ can be ignored for the theoretical study, however, it can determine the stability of two localization and is necessary to be observed by the energy dispersion. The non-Hermitian topology of $H_{k}$ with respect to the reference point $E_R$ is
equivalent to the Hermitian topology of the following doubled Hamiltonian~\cite{Lee2019,Masahito2018,Okugawa2019} with the basis vector $\tilde{v}=(\hat{a}_k,\hat{b}_k,\hat{a}_{k+2k_p},\hat{b}_{k-2k_p})^{\top}$ can be obtained in the supplementary materials where the open boundary condition is considered.
\begin{equation}
\tilde{H}_k=\left(
\begin{array}{cccc}
\left[\Delta' +2 J \cos (k)\right]\hat{\sigma}_z  & i gA^2 \hat{\sigma}_y  \\
 i gA^2\hat{\sigma}_y  & \Delta' \hat{\sigma}_z+h(k,k_p)\\
\end{array}
\right), \label{Hk} 
\end{equation}
where $h(k,k_p)=2J\left[\cos(k)\cos(2k_p)\hat{\sigma}_z-\sin(k)\sin(2k_p)\hat{I} \right]$ and $\hat{I}$ is the $2\times2$ identity matrix. The absolute value of the fluctuations (on top of the condensate) hopping from $u_n$ to $v_n$ will move to the $-k_p$ direction ($\hat{a}_k \hat{b}_{k-2k_p}^\dagger$ term ), meanwhile the polariton fluctuations hopping from $v_n$ to $u_n$ will move to the $+k_p$ direction ($\hat{b}_k \hat{a}_{k+2k_p}^\dagger$ term). Therefore, the absolute squre of the eigenstates of the BdG lattice are localized at both sides of the lattice as shown in Figs.~\ref{wf}(b)-(c).

The construction of the Hermitian Hamiltonian in the enlarged Hilbert space not only allows interpretation with the Bloch theory~\cite{skinmode1,Lee2019,Masahito2018,Okugawa2019}, but also highlights the differences of the four basis operators under the open boundary condition.  The boundary will let the momentum be cut off at the different values and cause the fluctuations to move to the edge and then be  localized. For example, the range of $k$ is $\left(-\pi,\pi\right)$, however, the range of $k\pm2k_p$ is $\left(-\pi\pm2k_p,\pi\pm2k_p\right)$. Some related work has been done in the SSH model with asymmetric hopping and interpreted with the GBZ theory~\cite{yokomizo2020nonbloch,Liu2020,yang2020nonperturbative}.


There are two EPs in the energy spectrum, so we need to calculate each winding separately. By unitary transition, the enlarged Hamiltonian can be written into a two blocks off-diagonal matrix and for each one we can define the winding around the reference energy $E_R$~\cite{skinmode1}

\begin{equation}
W_E=\frac{1}{2\pi i}\oint_{-\pi}^\pi \frac{d \log det\left[ H(k)-E_R I_{2\times2}\right]}{  dk}dk.\label{windingE}
\end{equation}
The total winding number is zero for the vanishing of $E$ and $E^*$ because the winding direction is completely opposite. However, we can get the integer winding number $\pm1$ or 0 in different topological regions. In our model, the reference energy is purely imaginary energies for the gapless $E$ and $-E^*$. The energies as a function of Hopping energies in the real space with OBC are shown in Figs.~\ref{real_BZ} (a) and (c). The critical points are labeled for the callapse of the imaginary energies.

\begin{figure}[h!]
\centering
\includegraphics[width=0.4\textwidth]{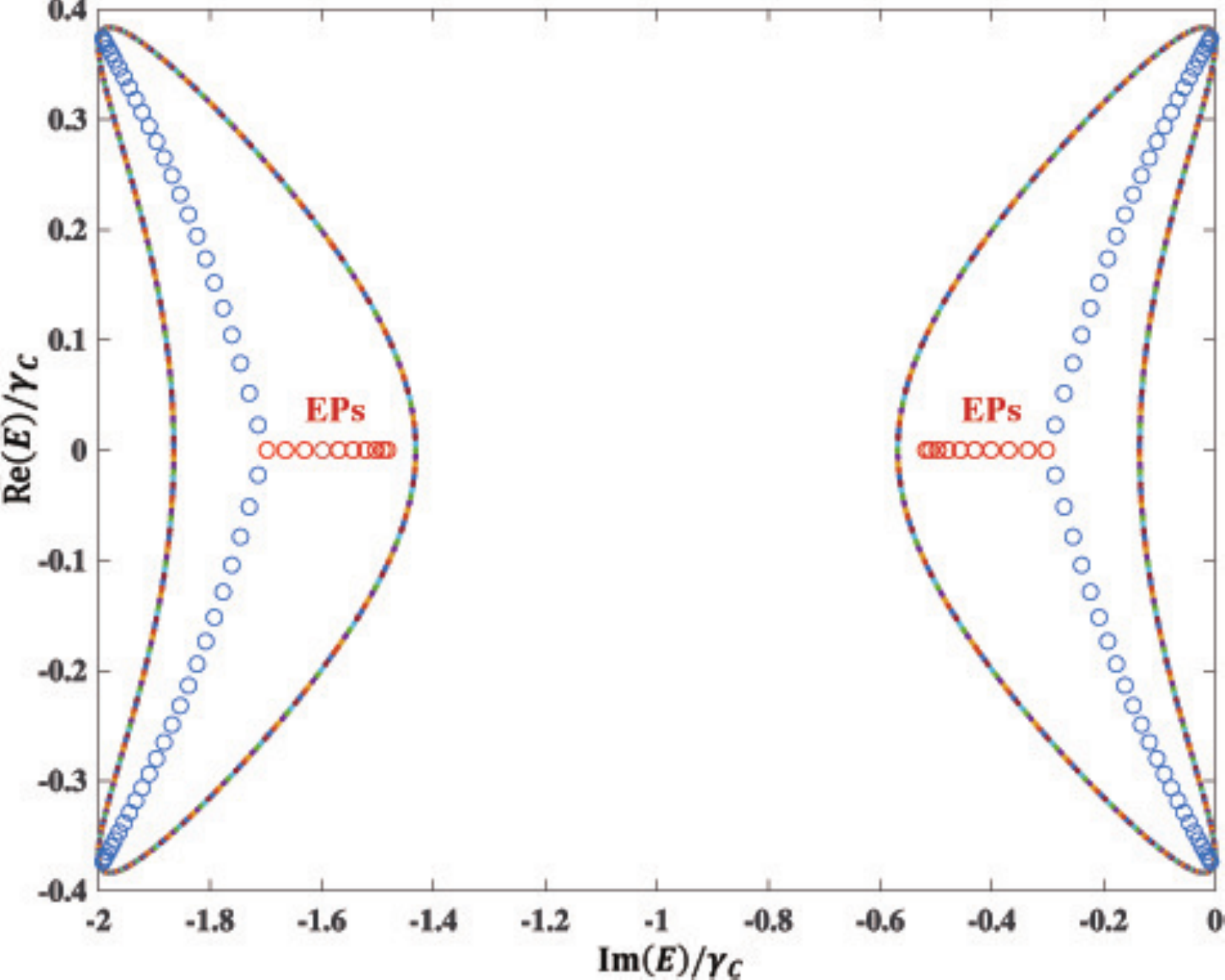}\\
\caption{ The real space energy profile with OBC (`circles') and the momentum space energy profile with PBC (`dots').  $J$=0.4 and other parameters are the same as in Fig. \ref{wf}. }\label{EPs}
\end{figure}

The different topological phases can be distinguished by the eigenenergy spectrum of  Eq.(\ref{Hk2}) and how many OBC energies in real space are encircled by the energy band in momentum space with PBC (see supplementary material):
\begin{itemize}
\item{{\emph{Nontrivial phase}} $0<\left|J\right|<\frac{\left(A^2+\Delta' \right) \sec (k_p)}{2} $: All eigenstates are localized at both sides and all OBC energies are circled by the PBC energies.   } 
\item{{\emph{Intermediate  phase}} $\frac{\left(A^2+\Delta' \right) \sec (k_p)}{2} \leqslant\left|J\right|<\frac{\left(A^2-\Delta' \right) \sec (k_p)}{2}$:  Parts of eigenstates are localized  and parts of OBC energies including EPs are encircled by the PBC energies.  } 
\item{{\emph{Trivial  phase}} $J=0$ or $J\geqslant \frac{\left(A^2-\Delta' \right) \sec (k_p)}{2}$: Parts or no eigenstates are localized and no EPs are encircled by the periodic boundary spectrum. } 
\end{itemize}
Here, the critical points are calculated by the phases of the eigenvalues of Eq. (\ref{Hk2}) and this result is corresponding with the winding number obtained by Eq. (\ref{windingE}) (see the supplementary materials). If the PBC energy dispersion has different linewidth (imaginary parts of  the energies) as a function of momentum, the skin modes will appear. Although the skin modes can appear in the above three phases,  the topological invariants are only $\pm1$ for two EPs in the nontrivial phase and the intermediate phase. The energies profile under the PBC can not encircle all the energies under the OBC in the intermediate phase. Therefore, all eigenstates are localized only in non-trivial phase and  the quantities of the localized eigenstates are corresponding with the quantities of OBC energies encircled by the PBC energies.

In the considered model all the calculations are done using the dimensionless variables. To get an idea about the physical parameters, we set $\gamma_C=0.1$ meV, which corresponds to a polariton lifetime around 3 ps. The coupling between the neighbouring micropillars becomes J=40$\mu$eV, which can be  tuned by adjusting the overlap between the micropillars \cite{Michaelis_de_Vasconcellos_2011}.  The non-linearity induced blueshift becomes 0.1 meV which is a routine observation in GaAs based samples~\cite{Estrecho2019}. Given the room temperature polariton micropillar chain has been demonstrated experimentally~\cite{Su_2020NP,Dusel_2020}, our scheme is also compatible with perovskite and organic samples. To make it more experimental friendly, the fluctuations can be made to have different polarization from that of the steady state by proper choice of the resonant driving and exciton-photon detuning~\cite{Sigurdsson2017,Mandal2019prb}.

\emph{Conclusion---}We consider the behaviour of fluctuations on a polariton mean-field stationary state in a one-dimensional lattice of coupled micropillars. By taking the stationary state as a plane wave, which can be resonantly injected with a suitably patterned optical field, we find that interactions in the system allow the presence of a bi-skin effect where fluctuations are localized at both edges of the lattice. Even though the underlying hopping of the system is Hermitian, the interactions allow an effective non-reciprocal coupling between particle-hole fluctuations. This results in a non-trivial topology, confirmed by calculations of the winding number.

\emph{Acknowledgements---} This work was supported by the Singaporean Ministry of Education, via the Tier 2 Academic Research Fund project MOE2018-T2-2-068.

\bibliography{mybib.bib} 
\clearpage
\title{  Interaction induced bi-skin effect in an exciton-polariton system}
\author{Xingran Xu}
\affiliation{Division of Physics and Applied Physics, School of Physical and Mathematical Sciences, Nanyang Technological University, Singapore 637371, Singapore}
\author{Huawen Xu}
\affiliation{Division of Physics and Applied Physics, School of Physical and Mathematical Sciences, Nanyang Technological University, Singapore 637371, Singapore}
\author{S. Mandal}
\affiliation{Division of Physics and Applied Physics, School of Physical and Mathematical Sciences, Nanyang Technological University, Singapore 637371, Singapore}
\author{R. Banerjee}
\affiliation{Division of Physics and Applied Physics, School of Physical and Mathematical Sciences, Nanyang Technological University, Singapore 637371, Singapore}
\author{Sanjib Ghosh}
\affiliation{Division of Physics and Applied Physics, School of Physical and Mathematical Sciences, Nanyang Technological University, Singapore 637371, Singapore}
\author{T.C.H. Liew}
\email{TimothyLiew@ntu.edu.sg}
\affiliation{Division of Physics and Applied Physics, School of Physical and Mathematical Sciences, Nanyang Technological University, Singapore 637371, Singapore}
\affiliation{MajuLab, International Joint Research Unit UMI 3654, CNRS, Universit\'e C\^ote d'Azur, Sorbonne Universit\'e, National University of Singapore, Nanyang Technological University, Singapore}
\maketitle

\begin{widetext}
\section{Stationary and excited states of the exciton polaritons under resonant pump}
In this section, we consider the behaviour of the Gross-Pitaevskii equation directly in the presence of fluctuations, which are modelled numerically. This will allow us to study the stability of the polariton mean-field solutions and access the fluctuation directly. We begin with the Gross-Pitaevskii equation for a driven-dissipative one dimensional micropillar chain, identical to Eq. (1) of the main text but accounting for an additional noise term ~\cite{Winkler2016}
\begin{eqnarray}
i\hbar \frac{\partial \psi_n}{\partial t}&=&\left(\Delta-i\gamma_C\right)\psi_n+ \sum_{<m>}J_{m,n}\psi_m
+g\left|\psi_n\right|^2\psi_n+F_n+i\hbar\frac{d\psi_{st}}{dt}.\label{HM}
\end{eqnarray}
Here, $\psi_{st}=2DdW$ accounts for the fluctuations
induced by white noise, where $dW$ is a Gaussian
random variable  and $D$ is the strength of the noise. The sum over $<m>$ represents summation over nearest neighbour sites. Other parameters are mentioned in the main text.  

\begin{figure}[h!]
\centering
\includegraphics[width=1\textwidth]{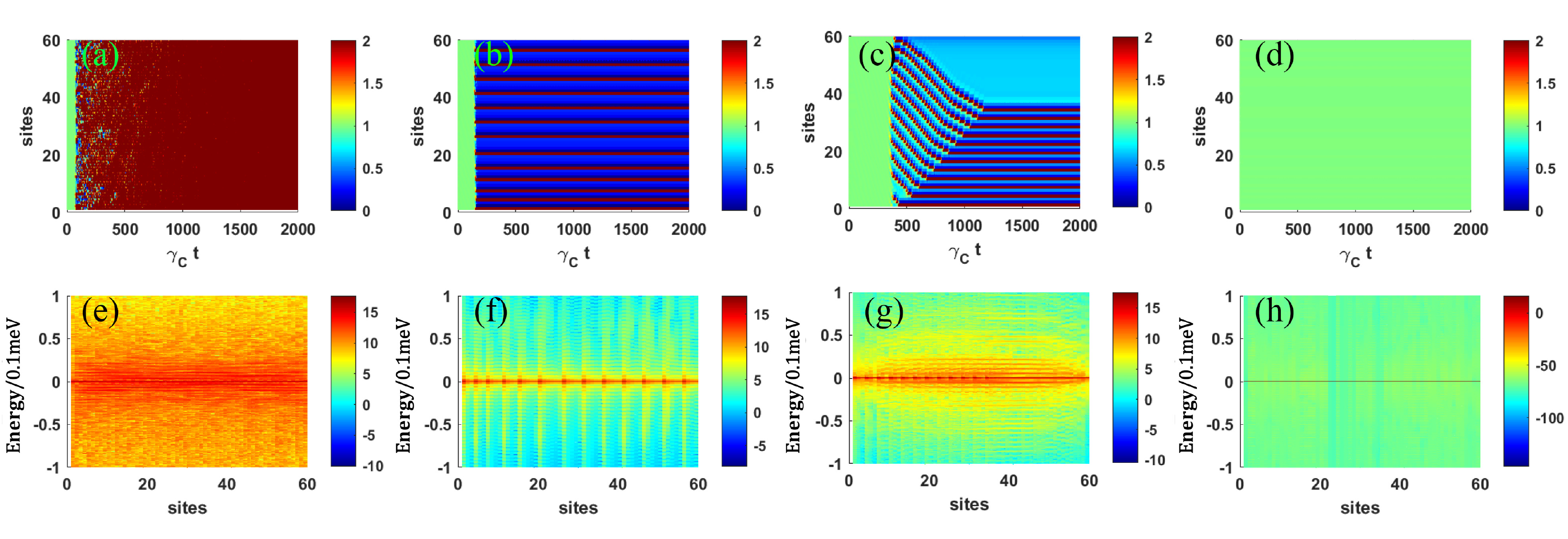}\\
\caption{The time evolution of the absolute square of the wavefunctions as a function of sites and  time $\left|\psi_{(n,t)}\right|^2$(the first row) or energies $\log_{10}\left[\left|\psi_{(n,\omega)}\right|^2 /\text{max}(\left|\psi_{(n,\omega)}\right|^2) \right]$ (the second row).  The parameters are used $\Delta$=-0.22meV, $J$=40$\mu$eV, $k_p$=$\pi/6$, $gA^2$=0.1meV, and $\gamma_C$=0.00meV, 0.05meV, 0.08meV, 0.10meV for different columns.}\label{timecrystal}
\end{figure}

As is shown in Fig. \ref{timecrystal}, we plot the time evolution of the wavefunctions of a finite-sized lattice with drive field in Eq. (2) and the boundary is considered in the main text. If there is no dissipation in the system, all fluctuations with different energies will be excited (see Fig.~\ref{timecrystal}(e)). However, if the dissipation gets larger, we can observe a time crystal like density distribution, where the intensity has a self-induced periodic oscillation and the parts of the energies get excited~\cite{Nalitov2019} (see Figs.~\ref{timecrystal}(b),(c),(f),(g)). If the dissipation is larger enough that can shift all imaginary parts of the energies below zero,  the system can have a stable plane-wave solution and only the zero energies modes.  
 \begin{figure}[h!]
\centering
\includegraphics[width=0.9\textwidth]{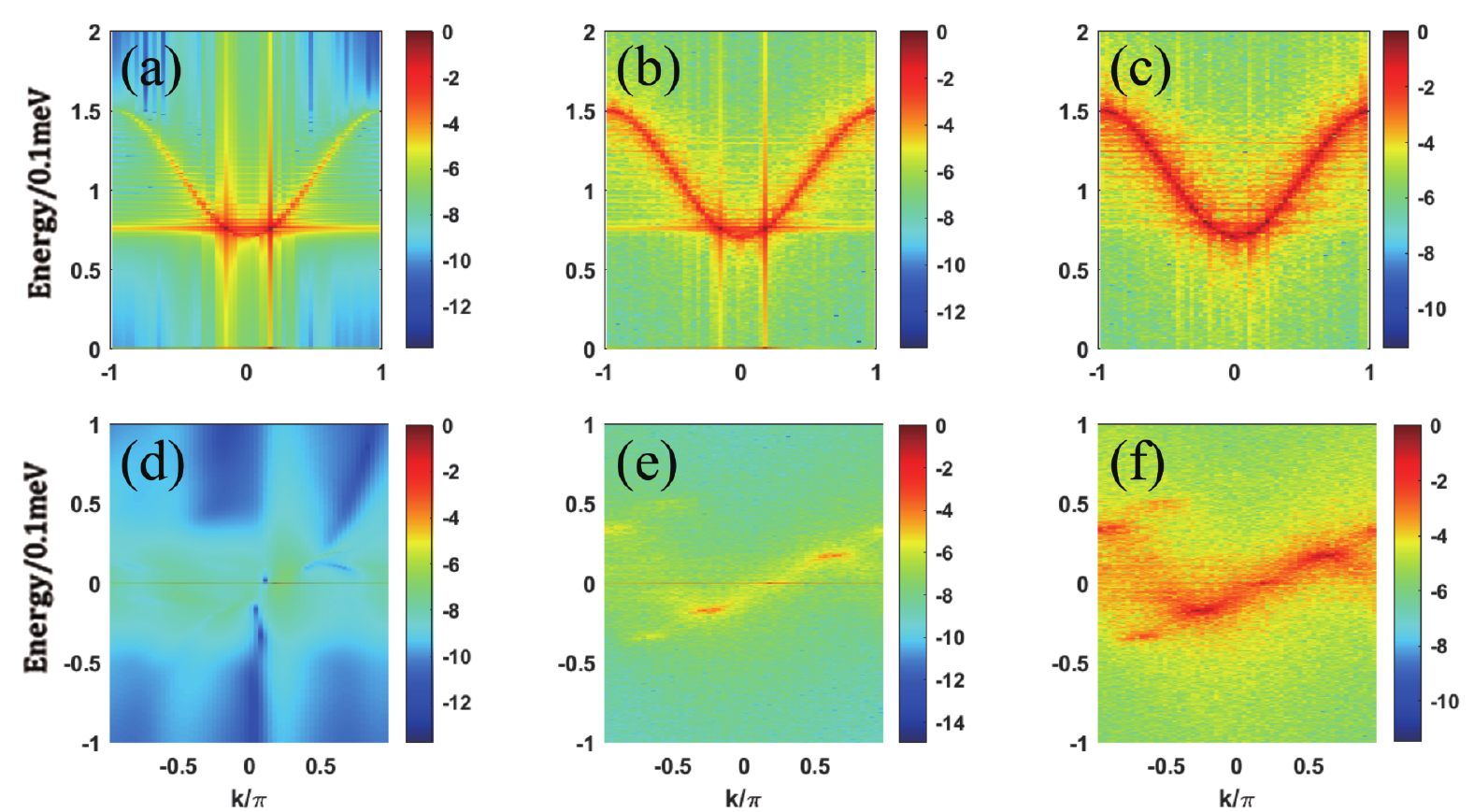}\\
\caption{ Energy dispersions for  the mean-field (the first column), the  mean-field with fluctuations (the second column), and only the fluctuations (the third column)  obtained from  Eq.(\ref{HM}) under the coherent pumping.  Parameters are used:  $\Delta$=-0.22meV, $J$=40$\mu$eV, $k_p$=$\pi/6$,  and without the nonlinearity the dissipation in the first row and $gA^2$=0.1meV,  $\gamma_C$=0.1meV in the second row. }\label{PO}
\end{figure}

As is shown in Fig. \ref{PO}(a)-(c), the energy dispersion of the free particle, in the absence of interactions, is symmetric at $k=0$, and the coherent pumping does not make the momentum shift but excites a strong signal of $k_p$ and $-k_p$ (because of the  scattering from the end of the chain). However, if we take the nonlinearity and the dissipation into account, the original energy dispersion is not stable as illustrated in Fig. \ref{PO}(d). If the fluctuations are added in Eq.~(\ref{HM}), we can see some obscure energy profile in Fig. \ref{PO}(e). In the third column of Fig. \ref{PO}, we plot $\left|\psi_n\left(k,\omega\right)-\psi_{(0,n)}\left(k,\omega\right)\right|$, which shows the energy dispersion of the fluctuations. The energy spectrum of the fluctuations under the periodic potential is still symmetric at $k=0$ and the linewidth is almost the same. The nonlinearity and the dissipation make energy dispersion of the fluctuations have a non-zero momentum ground state. 

\section{The topological invariants of the system}
In this section, we will calculate the eigenvalues and the winding number in detail. The OBC energies have two exceptional points (EPs) as is shown in Fig.~\ref{PhaseD}(b). We need to calculate each winding of them separately. The Hamiltonian in the real space is 
\begin{equation}
H_{n}	=\left(\Delta'-i\gamma_{C}\right)\sum_{n}u_{n}^{*}u_{n}-\left(\Delta'+i\gamma_{C}\right)\sum_{n}v_{n}^{*}v_{n}+J\sum_{n}\left(u_{n+1}^{*}u_{n}-v_{n+1}^{*}v_{n}+h.c.\right)
	+gA^{2}\sum_{n}\left(e^{2ik_{p}}u_{n}^{*}v_{n}-e^{-2ik_{p}}v_{n}^{*}u_{n}\right),
\end{equation}
where $u_n$ and $v_n$ are different Bogoliubov modes. If we just consider the nearest neighbouring  hopping and use the Fourier transformation $u_n \rightarrow \frac{1}{\sqrt{N}}\sum_k \hat{a}_k e^{ikn}$ and $v_n\rightarrow \frac{1}{\sqrt{N}}\sum_k \hat{b}_k e^{ikn}$ with $\hat{a}_k$ and $\hat{b}_k$  the annihilation operators for the momentum $k$, we can obtain the Hamiltonian with
\begin{eqnarray}
H=\Delta' \sum_k\left(  \hat{a}_k^{\dagger}\hat{a}_k- \hat{b}_k^{\dagger}\hat{b}_k \right)+gA^2\sum_k\left[\hat{a}_k^{\dagger}\hat{b}_{k-2k_p}-\hat{b}_k^{\dagger}\hat{a}_{k+2k_p}    \right]+2J\sum_k  \left(\hat{a}_k^{\dagger}\hat{a}_k-\hat{b}_k^{\dagger}\hat{b}_k \right)\cos k-i\gamma_C.\label{HR}
\end{eqnarray}
When the system is large enough and the periodic boundary condition is taken, we can get the $2\times2$ Hamiltonian with the vector $(\hat{a}_k,\hat{b}_{k-2k_p})$ as illustrated in the main text. To prove this, we can build a unit cell that contains  $l=4\pi/k_p$ inner sites  
\begin{eqnarray}
H(k)=\left( 
\begin{array}{cc}
H &  \text{diag}\left(A^2, \cdots, A^2e^{2ik_p(l-1)}\right)_{l\times l}\\
\text{diag}\left(-A^2, \cdots, -A^2e^{-2ik_p(l-1)}\right)_{l\times l}  & -H^*
\end{array}\right),
\end{eqnarray}
with $H=(\Delta'-i\gamma_C)\hat{I}_{l\times l}+\text{diag}\left(J,-1\right)_{l\times l}+\text{diag}\left(J,1\right)_{l\times l}$.  The matrix $\text{diag}\left(J,\pm1\right) $  are the upper and lower shift diagonal matrix with the matrix element $J$ and the Bloch theory gives $H_{1,l}=J e^{ik}$ and $H_{l,1}=J e^{-ik}$. The energy profile is corresponding with the $2\times2$ Hamiltonian  in the main text.

However, if we consider the finite size and assume the momentum $k$ changes continuously,  the effective Hamiltonian in the momentum space with vectors $v=(\hat{a}_k,\hat{b}_k,\hat{a}_{k+2k_p},\hat{b}_{k-2k_p})^{\top}$ ignoring the energy shift $\gamma_C$ can be written as
\begin{eqnarray}
\tilde{H}\left(k,k_p\right)=\left(
\begin{array}{cccc}
\Delta' +2 J \cos (k) & 0 & 0 & gA^2  \\
 0 & -\Delta' -2 J \cos (k) & -gA^2  & 0 \\
 0 & gA^2  & \Delta' +2 J \cos (k+2 k_p) & 0 \\
 -gA^2  & 0 & 0 & -\Delta' -2 J \cos (k-2 k_p) \\
\end{array}
\right). \label{Hks} 
\end{eqnarray} 
The Hamiltonian in the momentum space has the symmetry of $\tilde{H}(-k,-k_p)=\tilde{H}(k,k_p)$, $S _z \tilde{H}(k,k_p)S _z=\tilde{H}(k,k_p)^{\dagger}$, and $S _x \tilde{H}(k,k_p)S _x= \tilde{H}(k,k_p)$ with $S_{x,y,z}=I_{2\times2}\bigotimes \sigma_{x,y,z}$ with $\sigma_i$ is the Pauli matrix. The profile of the energy dispersion is highly symmetric and having  two EPs if the Hamiltonian is non-Hermitian with $A$ is a purely real number.

By diagonalization, the four eigenenergies can be obtained
\begin{equation}
E_{1,2}(k)=\pm\sqrt{f(k)}-2 J \sin (k_p) \sin (k-k_p),
\end{equation}
with 
\begin{equation}
f(k)=
-g^2A^4+\Delta'^2+J^2 \cos (2 k_p) \cos (2 k-2 k_p)+J^2+2 \Delta'  J \cos (k)+2 J \cos (k-2 k_p) (\Delta' +J \cos (k))  \label{fkn},
\end{equation}
and $E_{3,4}(k)$=$E_{1,2}(k+2k_p)$.  Meanwhile, the related wavefunctions are
\begin{eqnarray}
\ket{\psi_{1(2)}}=\left( \begin{array}{c}
-\frac{\Delta' -2 J \cos k_p \cos (k-k_p) \pm \sqrt{f(k)}}{\sqrt{2}A^2}\\
0\\
0\\
1/\sqrt{2}\\
\end{array} \right),  \ket{\psi_{3(4)}}=\left( \begin{array}{c}
0\\
-\frac{\Delta' -2 J \cos k_p \cos (k+k_p) \pm \sqrt{f(k+2k_p)}}{\sqrt{2}A^2}\\
1/\sqrt{2}\\
0\\
\end{array}\right),
\end{eqnarray}
where, the function $f(k)$ determine all the topological properties of the system.
 \begin{figure}[h!]
\centering
\includegraphics[width=0.9\textwidth]{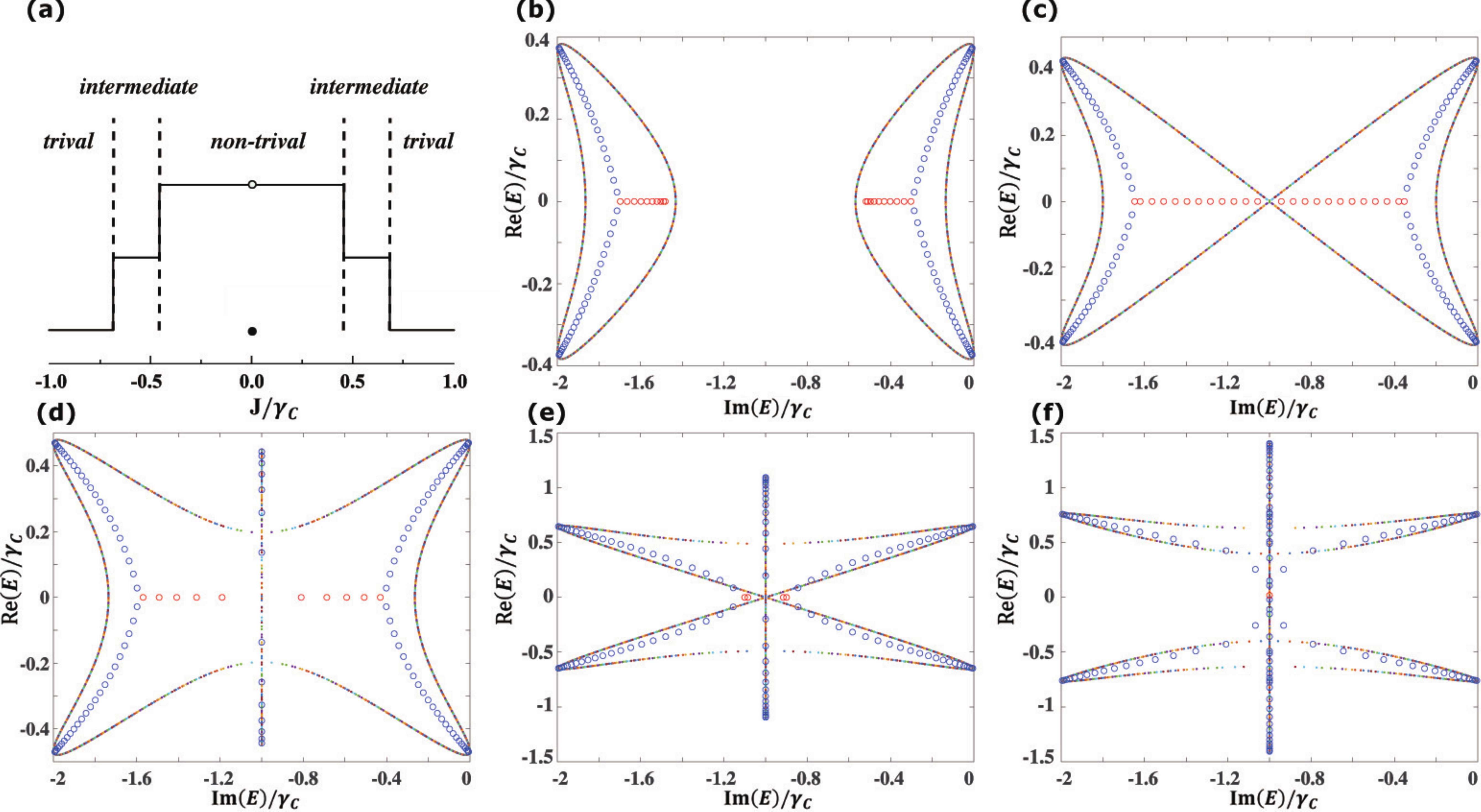}\\
\caption{(a) Phase transition diagram along with the change of the hopping energy $J$ and (b)-(f) the energy profiles obtained by OBC `dots'  and PBC `circles' .  Parameters are used:  $\Delta'$=-20$\mu$eV, $\gamma_C$=0.1meV, $gA^2$=0.1meV, $k_p$=$\pi/6$,  and $J$=40$\mu$eV, 46$\mu$eV, 60$\mu$eV, 69$\mu$eV, 80$\mu$eV for (b)-(f). }\label{PhaseD}
\end{figure}
The real energy will collapse at $k=\pm \pi/2$ for $E_{1,2}(k)=E_{3,4}(k)$, but the imaginary energy will have a jump across the EPs. To define each winding of the EPs, Eq.(\ref{Hks}) can be written into two blocks off-diagonal form
\begin{eqnarray}
\tilde{H}(k)&=&UH(k)U^{-1} \nonumber\\
&=&\left(
\begin{array}{cccc}
 0 & 0 & -\Delta' -2 J \cos (k) & -gA^2 \\
 0 & 0 & gA^2 & \Delta' +2 J \cos (k+2 k_p) \\
 -\Delta' -2 J \cos (k-2 k_p) & -gA^2 & 0 & 0 \\
 gA^2 & \Delta' +2 J \cos (k) & 0 & 0 \\
\end{array}
\right),
\end{eqnarray}
with the unitary transition $U=\left(
\begin{array}{cccc}
 0 & 1 & 0 & 0 \\
 0 & 0 & 1 & 0 \\
 0 & 0 & 0 & 1 \\
 1 & 0 & 0 & 0 \\
\end{array}
\right)$, and the winding number around two EPs can be calculated by 
\begin{eqnarray}
W_1&=&-\oint \frac{dk}{2\pi i} \partial_k\left \{ log\left[Det \left(
\begin{array}{cc}
 -\Delta' -2 J \cos (k) & -gA^2 \\
 gA^2 & \Delta' +2 J \cos (k+2 k_p) \\
\end{array}
\right)  
    -E_R \hat{I}   \right]  \right\},  \label{winding1}\\
W_2&=&-\oint \frac{dk}{2\pi i} \partial_k\left \{ log\left[Det 
\left(
\begin{array}{cc}
 -\Delta' -2 J \cos (k-2k_p ) & -gA^2 \\
 gA^2 & \Delta' +2 J \cos (k) \\
\end{array}
\right)
    -E_R^* \hat{I}\right]\right\},\label{winding2}
\end{eqnarray}
where $E_R$ and $E_R^*$ are two reference energies with open boundary condition(OBC) in real space. If we take the $loop$ from the whole Bloch zone, the total winding number is zero, but for each winding, we have $W_{1,2}=\pm1$. $E$ and $-E^*$ will give a positive winding number while the $E^*$ and $-E$ give the negative one. As is marked in Figs. \ref{PhaseD}(b)-(f) with red circles, all EPs are occurring at the pure imaginary energies. The eigenenergies that go around the left EPs can give the winding $1$ and $-1$ for the right EPs in Figs. \ref{PhaseD}(b)-(d). The total winding is always 0 for the vanishing of $E$ and $E^*$.

\section{Critical points of the topological transition }

In this section, we will calculate the critical points of the topological phase transition. As is defined by Eqs. (\ref{winding1})-(\ref{winding2}), two EPs have opposite windings. Based on the winding number and the localization of the wavefunction, we can define three phases: topological, intermediate, and trivial phase. 

\begin{figure}[h!]
\centering
\includegraphics[width=0.8\textwidth]{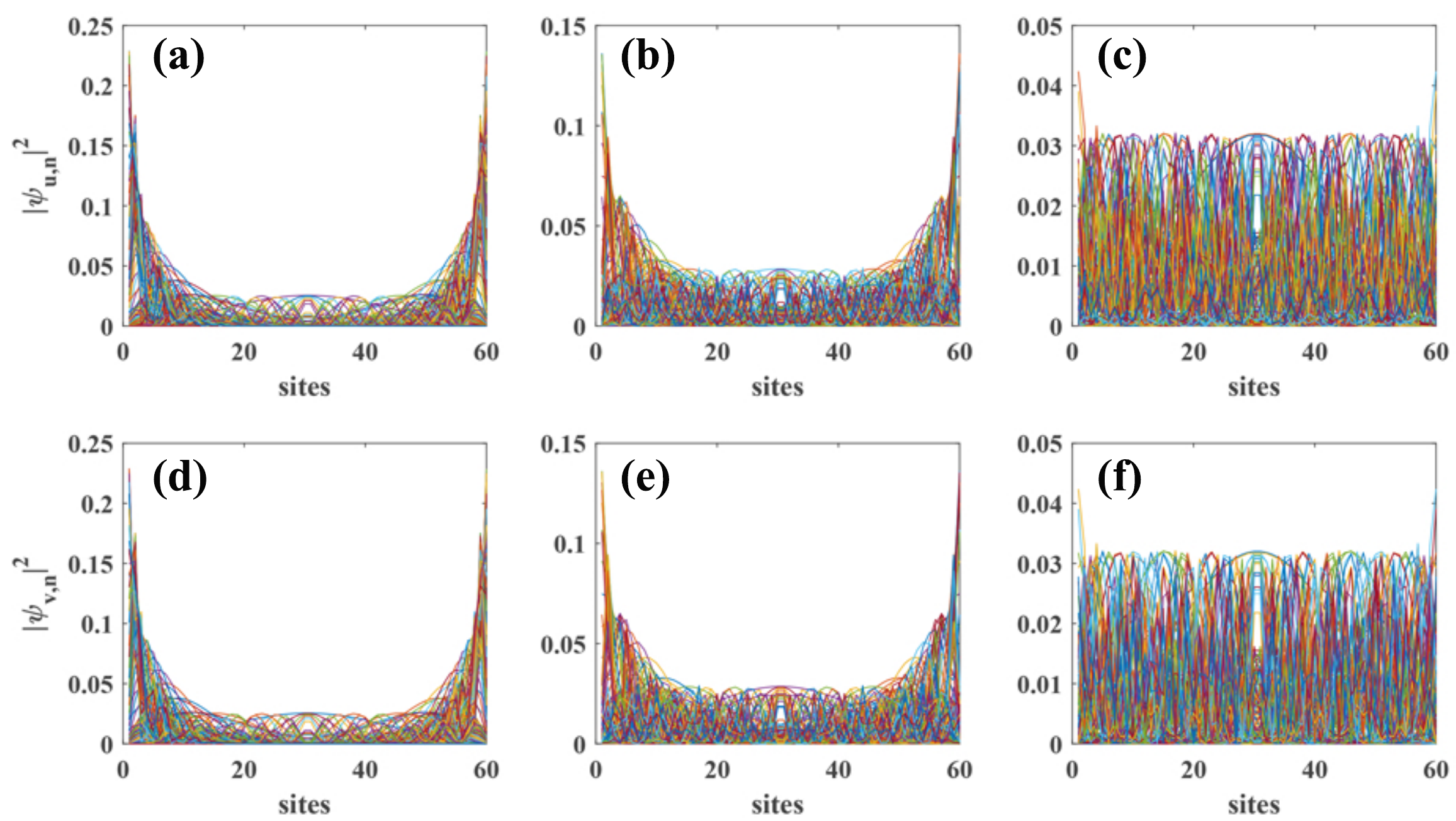}\\
\caption{The wavefuntions of $u_n$ (the first row) and $v_n$ (the second row) sites obtained by OBC. The parameters are used $\Delta'$=-20$\mu$eV, $gA^2$=0.1meV, $\gamma_C$=0.1meV,  $k_p$=$\pi/6$,  and $J$=60$\mu$eV (the first column),  80$\mu$eV(the second column), 0.2meV(the third column).}\label{supwf}
\end{figure}

The HN model \cite{HNmodel1,HNmodel2} can be easily theoretically realized by $H=J_1e^{ikx}+J_2 e^{-ikx}$ with $J_1$ and $J_2$ are asymmetric hoppings between  two nearest sites and the energy spectrum can be obtained by:
\begin{equation}
H=\left(J_1+J_2\right)\cos\left(kx\right)-i(J_2-J_1)\sin\left(kx\right)\label{HNmodel}.
\end{equation}
The skin modes will vanish for $J_2=J_1$ and appear for $J_2\neq J_1$. Actually, if the system has the same energy dispersion like Eq. (\ref{HNmodel}) where different directions of the momenta have gain and loss at the same time, the skin modes will appear even with the same hopping by inducing other terms into the Hamiltonian.

To realize the skin effect, the complex dispersion will be like 
\begin{equation}
E_k=C_1\sin(k+C_2),\label{EKSIN}
\end{equation}
with $C_i$ must be complex numbers. The complex number of $C_i$ ensures the condensates have gain in one direction and loss in another direction forming skin modes.

The derivative of Eq.~(\ref{fkn}) is
\begin{equation}
f'(k)=-4 J \cos (k_p) \sin (k-k_p) \left[\Delta' +2 J \cos (k_p) \cos (k-k_p)\right],
\end{equation}
with the zero points $k=k_p$ for the maximum and $k=\text{arcos}(-\frac{\Delta'}{2J\cos k_p})+k_p$ for the minimum.  If Eq. (\ref{fkn}) is a purely imaginary number in the whole Bloch zone, all eigenstates are localized. The critical points for the topological transition are $J_1= \left(gA^2+\Delta' \right) \sec (k_p)/2$  and $J_2= (gA^2 - \Delta') \sec(k_p)/2$ which is corresponding with the numerical results  $J_1\approx 0.46$ in Fig. \ref{PhaseD} (a) and (c) and $J_2\approx 0.69$ in Figs. \ref{PhaseD}(a) and (e). 

The phase can de defined in two ways, one is the appearance of the skin modes and the other one is the winding of the EPs energies. In the non-trivial phase,  $0<\left|J\right|<\frac{\left(gA^2+\Delta' \right) \sec (k_p)}{2} $ all eigenstates are localized and the PBC energies can encircle all the OBC energies including EPs as is shown in Fig. \ref{PhaseD} (b).    In the intermediate phase,  $\frac{\left(gA^2+\Delta' \right) \sec (k_p)}{2} <\left|J\right|<\frac{\left(gA^2-\Delta' \right) \sec (k_p)}{2} $ parts of  eigenstates are localized (see Figs.\ref{supwf} (a) and (d))  and the PBC energies can encircle parts of  the OBC energies concluding the EPs (see Fig. \ref{PhaseD} (d)). In the trivial phase, $J=0$ or $J\geqslant \frac{\left(gA^2-\Delta' \right) \sec (k_p)}{2}$, parts of (see Figs. \ref{supwf} (b) and (e)) or no eigenstates (see Figs. \ref{supwf} (c) and (f)) are localized and the PBC energies can encircle parts of  the OBC energies except the EPs (see  Figs. \ref{PhaseD} (e) and f). At the critical points $J=\pm J_1$ or $J=\pm J_2$, the PBC energy will cross at $E_R=0$, and the modes $-i\gamma_C$ arise as extended eigenstates that can't be encircled by the OBC energies.
\end{widetext}

\end{document}